# Tuning of polarisation sensitivity in closely-stacked trilayer InAs/GaAs quantum dots induced by overgrowth dynamics

Vittorianna Tasco[1,3,†], Muhammad Usman[2,3,‡], Milena De Giorgi[1], and Adriana Passaseo[1]

[1]National Nanotechnology Laboratory, Istituto Nanoscienze CNR, Via Arnesano, 73100 Lecce, Italy.
[2]Tyndall National Institute, Lee Maltings, Dyke Parade, Cork Ireland.
[3]*Equal contribution authors*
Corresponding Emails: [†]vittorianna.tasco@nano.cnr.it, [‡]usman@alumni.purdue.edu

**Abstract:**

Tailoring electronic and optical properties of self-assembled InAs quantum dots (QDs) is a critical limit for the design of several QD-based optoelectronic devices operating in the telecom frequency range. We describe how a fine control of the strain-induced surface kinetics during the growth of vertically-stacked multiple layers of QDs allow to engineer their self organization process. Most noticeably, the present study shows that the underlying strain field induced along a QD stack can be modulated and controlled by time-dependent intermixing and segregation effects occurring after capping with GaAs spacer. This leads to a drastic increase of TM/TE polarization ratio of emitted light, not accessible from the conventional growth parameters. Our detailed experimental measurements supported by comprehensive multi-million atom simulations of strain, electronic, and optical properties, provide in-depth analysis of the grown QD samples leading us to depict a clear picture on atomic scale phenomena affecting the proposed growth dynamics and consequent QD polarization response.

**Keywords:** Quantum dot stacks; Optoelectronics; Polarization response; Strain; Electronic Properties; TM/TE ratio;

# 1. Introduction

Since their theoretical proposal in 1981 by the seminal paper of Arakawa and Sakaki [1], semiconductor self-assembled quantum dots (QD) have inspired an extremely high quantity of experimental studies aimed at the employment of such nanostructures in a variety of optoelectronics devices, due to their intriguing atomic-like electronic properties. The flexibility in controlling their geometry by means of growth conditions and, in certain cases substrate patterning [2], has stimulated several studies on more complex structures able to add further potentialities to lens-shaped or pyramidal nanostructures commonly obtained by the Stranski-Krastanov process. In this context, one relevant physical example are closely stacked quantum dots, either consisting of QD layers separated by a thin GaAs spacer [3-5], or without using any GaAs spacer (also known as columnar QDs [6] or quantum posts [7]). In these kinds of nanostructures, the strong compressive biaxial strain component at the center of the typical flat shape dot can be reduced to zero or towards tensile values by increasing the stack height (adding QD layers), thus providing the optical polarization insensitivity desirable for relevant technological applications such as semiconductor optical amplifiers for high speed communication networks [8-10]. Such a requirement cannot be fulfilled by standard flat dome-shape QDs, where the biaxial compressive strain induces the valence band splitting into the heavy hole and the light hole states, thus providing strongly TE polarised optical transition [11]. This connection between quantum dot electronic structure and their morphology highlights the importance of complex and interplaying phenomena occurring during growth and overgrowth of this kind of nanostructures.

QD growth and capping procedure are lattice mismatched heteroepitaxial processes where strain release, segregation, faceting, intermixing and strain enhanced diffusion may occur at the island/cap interface [12, 13]. We recently demonstrated how the actual single layer QD composition profile is relevant in determining real optical behaviour of these InAs-based nanostructures [14]. This, in turn, is determined by complex nanoscale phenomena such as In-Ga intermixing and atomic segregation occurring during overgrowth with GaAs. In the case of multi-layer stacked QD devices, these phenomena become even more complex, due to the underlying strain field related to buried QDs, and the consequent QD shape/composition is more difficult to be assessed, also with the most actual and sensitive experimental investigation techniques. All these effects generate a complex scenario involving kinetics and thermodynamics but, at the same time, represent strategic tools for engineering the structural and optical properties which these nanostructures can exhibit.

So far, several QD engineering solutions have been explored [6, 15-18] for both InAs/GaAs and InAs/InP material systems, in order to change their optical properties by varying their aspect ratio and/or by controlling the biaxial strain in QDs thanks to strained barriers. The shape anisotropy has been engineered either embedding a single layer of QDs into InGaAs strain reducing layers [16, 17], or by closely stacked QDs (CSQDs) [15] and columnar QDs (CQDs) [10] formed by several repetitions of ML-thin InAs/GaAs layers. In particular, the last two QD geometries have shown a significant enhancement of TM-mode photoluminescence [3, 5, 8, 19] due to a reduction of the biaxial strain in the centre of the CQDs and CSQDs.

In this work, we show the engineering of shape anisotropy and related polarisation sensitivity of Stranski-Krastanov closely stacked InAs/GaAs QD layers by overgrowth phenomena and surface kinetics. The analysis reported is based on photoluminescence (PL) measurements at low temperature (10K) as a function of excitation power, and at room temperature as a function of polarization state, and atomic force microscope (AFM) investigations and is theoretically supported by multi-million atom calculations of strain, electronic structure, and optical spectra. Our results suggest that different QD reorganisation processes related to longer post-growth interruption time and consequent different surface strain condition can be an alternative tool to tune the actual polarisation behaviour in InAs nanostructures and to obtain TM/TE polarization ratio as high as 0.8.

## 2. Experimental Details:

The self-assembled QD structures used in this study were grown on semi-insulating GaAs substrates, by a COMPACT 21- Riber Molecular Beam Epitaxy (MBE) system equipped with reflection high energy electron diffraction (RHEED) gun to monitor *in-situ* the surface evolution during growth. After the growth of a GaAs buffer layer at 600°C, substrate temperature was lowered down to 500°C and QDs were formed by covering such a buffer with 2.8 MLs of InAs. The 2D-3D transition after deposition of ~1.7ML was demonstrated by the RHEED pattern evolution from streaky-like to spotty-like. Afterwards, dots were immediately capped by a GaAs spacer layer grown at the same low temperature. A single layer QD sample, labelled sample 1, was grown as reference, with a 20 nm GaAs cap terminating the structure. The stacked samples for polarisation investigation were realised by growing 3 QD layers closely spaced by 5 nm of GaAs. For all samples the final GaAs capping layer was 20 nm thick, whereas for the formation of QDs in the second and third layer only 2 ML of InAs were deposited, because the high strain field coming from the underlying QDs reduces the critical layer thickness of the stacked layers [12]. In our experiments, we focused on the evolution of QD shape and composition with respect to the growth interruption time (GIT) used before the deposition of the $2^{nd}$ and the $3^{rd}$ QD layer on the GaAs thin spacer. Such a step was always performed in As rich conditions, with a duration of 20 sec (labelled sample 2) and 120 sec (labelled sample 3). The resulting polarisation behaviour of the multi-stacked structures was found to be heavily affected and tuneable by this feature.

In order to experimentally study this effect, the stacked samples were lithographically processed to define apertures of 400 μm diameter inside a metal thin film (40 nm Ti/120 nm Au), thermally evaporated on the sample surface. The apertures were designed to optically pump a limited area of the samples, while the metal cover blocks the emission from the top of the sample, thus providing a consistent and reliable measurement of polarisation state of the edge emitted signal [20]. The PL signal excited from the top with an Ar+ laser (λ = 514 nm) collected from the [110] cleaved edge surface of the samples was filtered by a linear polarizer and detected by a NIR CCD positioned at the end of a 320 mm spectrometer.

## 3. Theoretical Models and Simulations:

In order to provide understanding of experimental measurements, we performed multi-million atom simulations of electronic and optical properties of the grown QD samples and compared calculations with the experimental results. The simulations were performed using well-known NanoElectronic MOdeling (NEMO 3-D) simulator [21, 22], in which the strain is computed from atomistic Valence Force Field (VFF) model [23-25] and the electronic structure is computed by solving twenty-band $sp^3d^5s^*$ tight-binding Hamiltonian [26]. The polarization dependent interband optical transition strengths (TE and TM modes) are calculated using Fermi's Golden Rule [3, 27]. Furthermore, the calculated polarization dependent TE and TM modes are cumulative sum of the inter-band optical transition strengths between the lowest conduction band state and the highest five valence band states, where each transition is broadened by multiplying with a Gaussian function of FWHM=5 meV and its peak centred at the energy of transition [3], to take into account the effect of inhomogeneous broadening stemming from small variations of QD dimensions, etc.

It should be noted here that theoretical modelling of QD stacks possesses a two-fold challenge. Firstly, it requires modelling techniques with atomistic resolution that can calculate electronic structure and optical properties including correct symmetry and interface roughness. Secondly, the relatively large size of QD stacks requires calculations to be performed over several millions of atoms to properly include the long-range effects of strain. Our atomistic simulations performed over about 18 million atoms in the device take both factors into account and therefore are capable of explaining experimental data with a high degree of accuracy, thereby allowing us to identify the main atomic-scale mechanism below the observed polarization response of investigated samples.

## 4. Experimental Results and Discussions

The present analysis starts from our recent study [14], where we described the dependence of the polarization-dependent optical emission on the actual chemical composition of QDs for a single QD layer, as modulated by atomic scale phenomena occurring during overgrowth, *i.e.* In-Ga intermixing, strain-enhanced diffusion, and In segregation. The effect of these phenomena has been accurately described only with a complex QD compositional model (shown in the left inset of figure 1) having an In-rich core region surrounded by a lower In content external shell, which actually allows to fit the experimental data in terms of both optical emission (PL peak wavelength) and polarization response (TM/TE ratio), as shown in figure 1.

A more complex atomic scale dynamics is expected for a stacked QD layer configuration, where the buried QD layers strongly modify subsequent QD growth with the strongly localized strain field originating at each QD apex [12, 13]. In this scenario all the involved growth parameters, and among them the GIT can play a crucial role in the assessment of actual QD structure, stoichiometry, and resulting optical polarization behavior.

## 4.1 Impact of GIT on QD stack morphology and optical properties

InAs QDs undergoes several atomic scale phenomena when overgrown with GaAs cap layers [28, 29]. The evolution of these phenomena can be correlated to the morphology of GaAs capping surface, as shown in Ref. [28]. Therefore, in order to obtain a realistic picture of the morphology of the grown QD samples, we performed AFM analysis on the top GaAs surface of samples 1, 2 and 3 respectively, (shown in figure 2). As expected, the underlying QDs lead to a mound-like morphology in the top surface of all the samples. In the reference single QD layer (sample 1), a mound density of $4 \times 10^{10}$ cm$^{-2}$ is measured, with average height and diameter of 1.5 nm and 50 nm, respectively (figure 2(a)). Such morphology directly reflects the dot size and distribution observed in uncapped single QD layer sample (figure 2(d)).

A significant change is observed as the interruption time (GIT) is increased from 20s (sample 2) to 120s (sample 3) in the two stacked samples. In sample 2, mounds are rather well organised, forming oriented and elongated chains, with average length of hundred nm and average lateral size of 30-40 nm (figure 2(b)). Elongation along [-110] direction is usually observed during initial QD overgrowth by GaAs or other compounds [28, 30, 31], as an effect of anisotropic surface diffusion, since Ga adatoms tend to migrate far away from the high stress regions. After few monolayers, such an elongation is usually no more evident, rather leading to an undulated GaAs surface morphology prevailing up to capping thickness of the order of few tens of nanometers [28]. Persistence of elongated morphology after 20 nm GaAs capping in sample 2 suggests that buried stacked QDs do not have an ideal cylindrical symmetry. Conversely in sample 3, the overgrown GaAs morphology (figure 2(c)) is very similar to the one of the single layer sample (figure 2(a)) apart from an increase of mound lateral size up to 60 nm. Moreover, for this long GIT an uncapped three QD layer sample was also grown for comparison. The AFM analysis on such a sample (figure 2(e) and related diameter histogram) clearly shows that QD size distribution and density reproduce the ones of the single layer (figure2(d) and related diameter histogram).

PL spectra collected at 10K for sample 1, sample 2 and sample 3 show that the QD ground state (GS) emission blue-shifts in both the stacked samples with respect to the single layer sample. A line-width shrinking of the GS spectrum with the stacking is also observed: from 31 meV in the single layer to 24 meV in sample 2 and to 19 meV in sample 3, suggesting a wide size distribution in the reference sample, whereas a size filtering effect occurs among the vertically coupled QD layers [32].

The blue shift of the GS emission with the stacking at thin spacers (in strong quantum coupling regime) with respect to single QD layers, was found in previous studies and was generally attributed either to strain driven In-Ga intermixing [15, 33, 34] or to a reduced wave function overlapping along the stacked layers [35]. Actually, the variation of the GIT parameter in our growth can promote the time-dependent In out diffusion phenomenon from the buried QDs. This, in turn, modifies the nucleation dynamic of subsequently grown QD layers, since the exposed surface conditions are consequently altered. Therefore, different localization effects of electron or electron/hole wave functions can be expected in the final QD stack, finally modifying the resulting QD polarization response and enhancing the TM/TE ratio.

By the analysis of low temperature PL spectra as a function of optical pumping power density (figure 3), performed by multi Gaussian curve fitting, it can be noted that the sample 1 (single QD layer) PL spectrum exhibits (figure 3 (a)) a clear presence of the ground state (GS) and the first excited state (ES) as a function of excitation power, separated by large energy spacing (~80 meV), typical for flat single QD layers [36]. Similarly, the PL spectrum of sample 2 (figure 3 (b)) presents a second peak at the high energy side, spaced by 58 meV from the GS, and exhibiting the band filling behaviour with increasing power density expected from a first excited state. On the other hand, in sample 3 (figure 3 (c)) a shoulder is evident at the high energy side, at about 21 meV from the GS emission. The relative intensities of the two lowest energy peaks were found to be not dependent on the excitation density. Therefore, we associated this additional emission to the anti-bonding state emission due to the QD molecule formation [33, 37, 38], demonstrating stronger wave function hybridization in this sample with respect to sample 2.

The most striking effect of employing a different GIT is highlighted in the polarization response of the two stacked samples. Figure 4 shows the measured polarized RT PL spectra of sample 2 and sample 3, taken under the same experimental conditions. While the single QD layer sample is strongly TE polarised (TM/TE = 0.26) as shown in figure 1, the polarisation ratio is progressively increased in sample 2 ($\rho$ = 0.62) and in sample 3 ($\rho$ = 0.8) where longer growth interruption times are employed. This demonstrates the possibility of tuning the polarization response towards isotropic value (TM/TE ~ 1.0) from QD nanostructures by controlling the overgrowth process, without growing a large number of stacked QD layers which can increase the overall strain of the structure.

## 4.2 Discussions

The QD formation and their GaAs overgrowth, at such a high substrate temperature (510°C), are well known [13, 28-31] to result from the complex interplay between In-Ga intermixing at the dot/barrier interface, In vertical segregation through the dot height and In desorption out of the dot and back to the vapor phase. Considering the observed AFM morphology ( figure 2 (d)) and previously performed TEM investigations [33], buried QDs of the first layer are probably left partially uncapped by the overgrowth with 5 nm GaAs spacer. At this stage, different growth interruption times after the spacer growth and before further InAs coverage can cause a modification of strain-induced surface conditions, thus altering subsequent QD formation.

When the growth interruption is delayed after capping, it is likely that, both, In-Ga intermixing inside the buried nanostructures is enhanced, and In atoms from the partially uncapped apex of buried dots tend to out-desorb in a relatively large, time dependent amount. Therefore, the stress induced on the reconstructed surface by the underneath lattice mismatch is lowered and during the overgrowth with second QD layer, the migration length of impinging In ad-atoms is consequently increased. In general, strain-driven migration leads In ad-atoms to diffuse towards the buried QD positions which represent favorable nucleation centers for second and third QD layer, in both GIT cases, giving the typical vertically self-organized growth behavior, as confirmed by the shrinkage of PL line width with the stacking. In other words, in both stacked samples, buried QDs act as nucleation sites for second and third QD layer. The key point is that in the case of sample 2 with short GIT (20s), the time-dependent In-Ga intermixing in the buried dots slightly reduces the underlying strain field, and a short surface migration length of In adatoms is expected. In this case stacked QDs are enlarged in size with a low degree of cylindrical symmetry. Indeed, the AFM image of GaAs cap surface over the 3 QD layers shows an elongation enhanced during the overgrowth. Moreover, in sample 2 the increase in lateral size along the stack is also reflected in the weaker overlapping of electron wave functions and the absence of bonding and anti-bonding transitions in the luminescence spectra.

For a longer interruption time, as for sample 3, increased In out-diffusion in the buried dots produces nucleation sites with a lower binding energy for the In ad-atoms provided for the second and third layer growth. This leads to a QD formation dynamics more similar to that one of the first QD layer with respect to the lower GIT case, resulting in stacked QDs with morphology closer to the buried ones (dome shaped). In this case, stacked QDs should be characterized by a more pronounced cylindrical symmetry, as confirmed by the AFM results of the 3$^{rd}$ uncapped layer that also exhibits the same size distribution and density of first QD layer ( figure2(d) and figure 2(e)). Such a shape and size preservation, in turn, leads to the reduced linewidth broadening of sample 3 (with respect to sample 2) and reduced low temperature blue shift PL emission (with respect to single layer). Moreover, "columnar" shape of stacked QDs in sample 3 gives the observed presence of bonding and anti-bonding states in sample 3 ($GS_b$ and $GS_a$), as evident from figure 3(c).

## 5. Numerical Analysis based on Atomistic Simulations

A better understanding of the proposed dynamics can be obtained by exploiting multi-million atom simulations that allow us to correlate dot shape/morphology and consequent optical and electronic properties, with a particular focus on strain field and polarization sensitivity.

## 5.1 Theoretical modelling of QD stack confined wave functions as a function of GIT

Firstly, we performed our atomistic simulations to evaluate the impact on the wave function confinement of the different QD evolution along the stack related to the employed GIT. For this purpose, we simulated the following two QD systems, based on the analysis of AFM investigation:

**MS1:** A stack composed by three vertically stacked layers of dome-shaped QDs, with increasing diameter and decreasing height: the lowest QD has diameter of 15 nm and height of 5 nm (as in Ref. 14), the middle QD has diameter of 20 nm and height of 4 nm, and the top QD has diameter of 25 nm and height of 3 nm.

**MS2:** A stack comprised of three identical vertically stacked dome-shaped QDs (as observed in sample 3 with GIT 120s), with equal diameters of 15 nm and heights of 5 nm.

In both cases, MS1 and MS2, all the QDs are placed on top of 0.5 nm thick InAs wetting layers.

Figure 5 gives an insight on the evaluated electron confined states (E1, E2 and E3) for the two proposed QD systems (MS1 and MS2). In this figure electron wave function spatial distribution and symmetry can be visualized by side and top view maps. Energy difference values between electron states (E2-E1 and E3-E2) are also indicated for the investigated configurations.

For the trilayer QD stack with GIT=120s (MS2) shown in figure 5(a), the QD layers are strongly coupled and, therefore, the lowest two states E1 and E2 show the formation of bonding and anti-bonding states, respectively, separated by relatively smaller energy difference of 22 meV. In this system, the first excited (p) state is the third state E3 which is further separated by 40 meV from E2. This is in agreement with what we found from excitation dependent PL measurements for sample 3 (figure 3 (c)) where bonding and anti-bonding type transitions are observed separated only by 21 meV, validating the assumption of shape preservation and strong wave function localization along the QD vertical stack.

The hybridization of wave functions becomes significantly weak when the QD base diameter is assumed to increase along the vertical direction (MS1) as in the case of sample 2 (figure 5(b)). For this stack, our calculations indicate that E1 (s-state) is weakly hybridized and predominantly in the upper QD region. The next state E2 is the first excited state of p-type symmetry and is separated by 38 meV from E1, while the E3 state is the second s-state separated by only 4 meV from E2. This finding is in agreement with both excitation dependent, low temperature PL measurements (figure 3 (c)) and elongated morphology evidenced by AFM analysis shown in figure 2 (b). The two proposed stacking models (MS1 and MS2) are characterized by different strain profile, as shown in figure 6 where we plot line scans of hydrostatic ($\varepsilon_{xx}+ \varepsilon_{yy}+ \varepsilon_{zz}$) and biaxial ($\varepsilon_{xx}+ \varepsilon_{yy} - 2\varepsilon_{zz}$) strain components through the center of both trilayer stacks. Figure 6 (a) compares the hydrostatic components, clearly showing a decrease in the magnitude of the hydrostatic strain for the larger upper QDs (QD2 and QD3) in case of MS1 (sample 2). This will result in deeper conduction band edges for these QDs [39] and therefore the electron wave function will largely reside inside the larger QDs towards the top of the stack, in consistency with the wave function plots of figure 5 (b). The strong electronic coupling of QDs for the MS2 case accompanied with roughly equal magnitudes of the hydrostatic strain in the three QDs (due to their equal size) lead to strong hybridization of the electron wave functions forming bonding and anti-bonding states.

Figure 6(b) compares the biaxial strain components for the two trilayer stacks MS1 and MS2 indicating an opposite trend: the larger size of QDs in MS1 leads to an increase in the magnitude of the biaxial strain component (as shown by arrows in the figure 6(b)). This will result in larger splitting between the heavy-hole (HH) and the light-hole (LH) band edges for the larger QDs [39], thereby reducing the TM-mode component. The equal size of QD layers for MS2 favors the hole wave functions to reside inside the bottom most QD layer. Finally, a much reduced biaxial strain for MS2 (GIT=120s case) as compared to MS1 (GIT=20s case) is expected to significantly impact on the polarization-dependent optical properties of this stack.

## 5.2 Theoretical modelling of electronic and optical properties of long GIT stacked QDs

In order to quantitatively explain the observed increase in the TM/TE ratio for the sample 3, we performed simulations to calculate TM and TE modes for the vertical QD stack with the three identically sized QDs (15 nm diameter and 5 nm height). Since any tentative to simulate the overall optical response of the structure by considering QD with uniform composition, failed to explain the experimental measurements [14], the expected In/Ga intermixing effect is mimicked by using the two-composition model with a high In composition QD core surrounded by a lower In composition shell. We assigned different compositions to each region of the QDs as shown by the schematic of figure 7. The sizes of high In core region are also kept fixed to 11 nm height and 4 nm diameter in accordance with Ref. [14].

Our analysis of the single QD layer predicted the inner core of QDs with In content close to 1.0 and the outer shell with lower In content ($\leq$ 0.4) [14]. Therefore, we performed a set of systematic simulations as a function of QD compositions while keeping x1, x3, x5 $\geq$ 0.8 and x2, x4, x6 $\leq$ 0.4, and compared our computed values of the GS and TM/TE ratio with the corresponding experimentally measured ones.

The preliminary assumption was that each QD layer had identical composition of both, outer shell and inner core. The considered compositions are listed in table 1 (double compositions 1 to 6). For an outer shell In composition of 0.4 and an inner core In composition of 1.0, as found for the single layer sample [14] the values of GS and TM/TE ratio are 0.98 eV and 1.047, respectively, thus not fitting the measured values (1.057 eV and 0.8, respectively). The agreement between theory and experimental results can be improved by lowering the inner core composition to 0.9 and 0.8, that blue shifts the GS to 1.07 eV and decreases the TM/TE ratio to 0.88. Additional modifications of this assumption to get the TM/TE ratio closer to its experimental value, such as unrealistically small values of In composition in the core region obviously increases the GS energy. On the other side, any decrease in the outer shell composition (double compositions 4 to 6 in table 1) to blue shift the GS value are accompanied by very large values of TM/TE ratio, as expected by the resulting reduced strain field in the nanostructures.

Therefore, also based on the fact that the localized strain induced by the lower QD layers should lead to reduced In out diffusion in the subsequent QD layers, we have introduced a vertical gradient in the composition profiles of the QD stack. Firstly, the gradient is considered only in the outer-shells, keeping the inner core compositions identical for the three QDs, *i.e.* x2$\neq$x4$\neq$x6 and x1=x3=x5. The considered outer-shell compositions decrease from x2=0.4 down to x6=0.2 with a gradient $\Delta$x=0.1 (double compositions 7 to 10 in table 1). The calculation was repeated for inner core compositions (x1=x3=x5) varied from 1.0 to 0.8. A systematic increase in the value of GS energy from 0.98 eV to 1.07 eV and a corresponding decrease in the value of TM/TE ratio from 0.92 to 0.76 is then calculated for compositions 7 to 10.

As a second step, we introduced a composition gradient also in the inner core of QDs, while keeping the outer shell composition gradient of 0.1. The long GIT employed in sample 3 is likely to enhance In out diffusion and vertical segregation in buried QDs and, consequently, a vertically increasing gradient could be reasonably expected for the inner core composition. The corresponding simulated composition profiles are listed in table 1 (double compositions 11 to 14). A systematic increase of the GS energy along with the lowering of the TM/TE ratio was calculated as the average inner core composition is lowered from 0.95 to 0.85, with the best fit with experimental measurements found for the composition models 13 and 14.

The three double composition models that provided best match with the experimental data (lines 9, 13 and 14 in table 1) were also used to recalculate the electron wave function spatial distributions and symmetries (as shown in figure 8). In all these three cases, the presence of bonding and anti-bonding states, consistently with experimental PL (figure 3(c)), is confirmed, in agreement with what calculated earlier using pure InAs composition profile (figure 4(a)). The presence of bonding and anti-bonding states in sample 3 can then be inferred to the strong coupling between identically sized QD layers inside the stack. Furthermore, we also calculate and plot the highest three hole wave functions (H1, H2, and H3) in figure 9 for the three double composition models under study. All the hole wave functions are found to be confined in the bottom most QD layer. We also note from the comparison of the top views of the hole wave functions that the variations in the In/Ga intermixing only result in very small changes in the orientations of the hole wave functions.

Overall, we conclude that the In/Ga intermixing effect, regulated by the surface kinetics induced by underlying strain field, does only slightly affect the actual confinement and symmetry of the electron and hole wave functions in such a QD stack, whereas it is the crucial parameter impacting the related polarization response, giving us the possibility to obtain high TM/TE ratio even with only a triple QD layer stack.

## 5.3 Theoretical modelling of electronic and optical properties of short GIT stacked QDs

Based on our analysis of the AFM images (figure 2) and the electronic state simulations (figure 5), we have shown that the short GIT employed in sample 2 is likely to induce a lateral size increase in the stacked QDs, as commonly observed for closely spaced QD stacks. The model MS1 has already shown a reduction of the hydrostatic strain and a reinforcement of the biaxial strain towards stack top (figure 6). The TM mode intensity is therefore expected to decrease for the GIT=20s sample, as it is directly related to the intermixing of HH and LH character in the valence band states which reduces due to the larger magnitude of the biaxial strain thereby increasing the separation between the HH and LH bands [36]. In order to estimate only the impact of the QD lateral size along the stack (thus ignoring In/Ga intermixing effect), we applied atomistic calculations to calculate TM/TE ratios for the two model systems presented above (MS1 and MS2). We found a TM/TE ratio of only 0.48 for MS1 and of 0.84 for MS2. This difference is in agreement with the trend of polarization dependent PL measurements shown in figure 4, where TM/TE ratio for sample 2 is found much smaller (0.62) as compared to the value of 0.8 for sample 3. Therefore, we conclude that the lateral size of the QD layers inside the stack, controlled by the strain-induced surface dynamics, has a very strong impact on the polarization ratio.

Furthermore, since from our AFM analysis we would expect a much weaker In/Ga intermixing for sample 2, we again applied the two-composition model shown by schematic diagram of figure 7. Due to the increasing lateral size of the QD layers in sample 2, we proportionately increased the size of the inner core. Based on our understanding of the much reduced intermixing for the sample 2, we assume 5% In composition in the outer-shell region, *i.e.* $x2=x4=x6=0.05$ while the inner core compositions ($x1=x3=x5$) are varied from 1.0 to 0.9. A corresponding systematic increase in the peak GS energy from 0.95 eV to 1.01 eV with a TM/TE ratio ranging from 0.55 to 0.62 were found, in a good quantitative agreement with the experimental findings. This behavior indicates that the vertical composition profile along the stack has a weaker effect in sample 2, whereas QD shape play the dominant role in controlling the spatial distribution of wave functions and their overlaps.

## 6. Conclusions

In this work we describe how the growth of vertically stacked QDs is influenced by atomic scale phenomena of intermixing, segregation, and out-diffusion, affecting the surface kinetics of further provided In adatoms. The proposed dynamics were investigated by combining experimental investigation of QD morphological and optical properties with atomistic numerical calculations depicting a clear figure of wave function localization and strain field distribution along the stacked structure. The QD structural and optical properties were found to be affected by the interruption time after buried QD capping with GaAs which plays a relevant role in determining the strain energy conditions of surface exposed to stacked QD layers. As a result, the QD self organization process allows to achieve a strong increase of TM/TE polarization ratio of emitted light by only three closely stacked QD layers, promising for technological applications based on QD optoelectronic devices.

## Acknowledgements:

VT, AP, and MDG acknowledge the technical help of I. Tarantini and P. Cazzato. MU acknowledges the use of computational resources from http://nanohub.org operated by Network for Computational Nanotechnology (NCN) at Purdue University, USA.

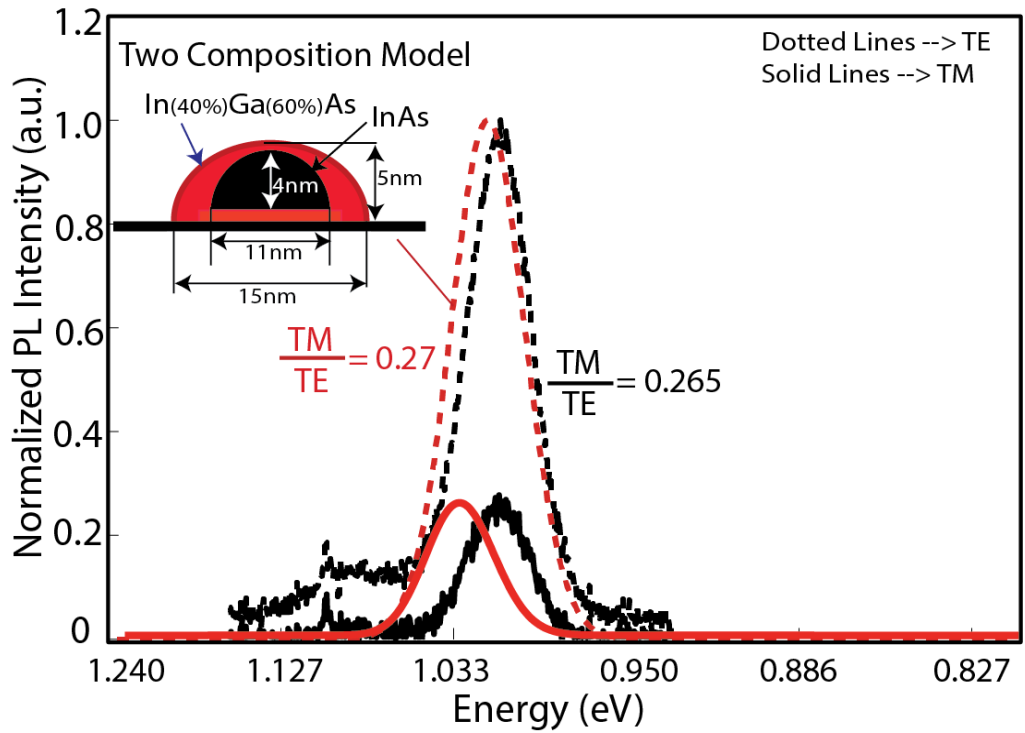

Figure 1: Comparison of calculated (red curves) polarization dependent optical spectra (TE and TM modes) with the experimentally measured (black curves) spectra for the single QD layer (sample 1). The insight on left shows the schematic diagram for the modeled two-composition scheme to mimic In-Ga interdiffusion.

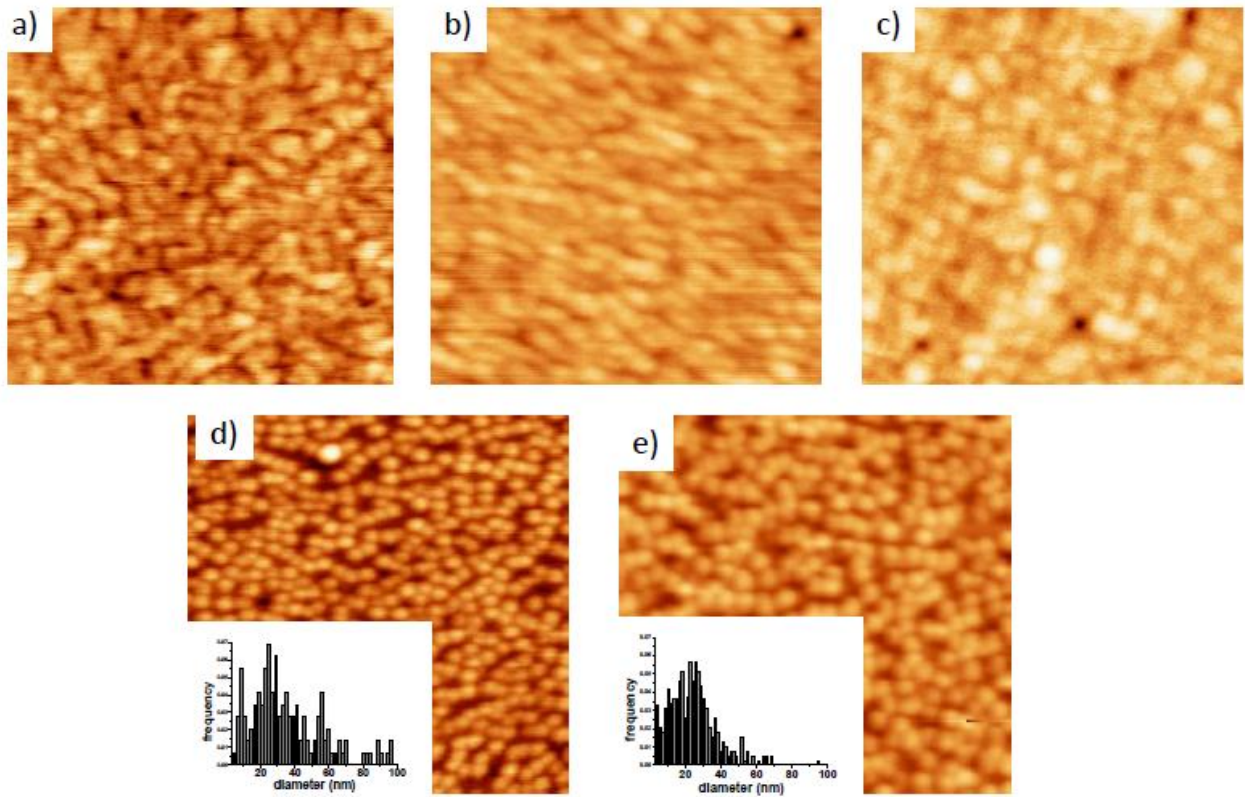

Figure 2: 1µmx1µm AFM scan images of for (a) GaAs surface of sample 1 (capped single QD layer), (b) GaAs surface of sample 2 (triple QD stack grown with 20s GIT), and (c) sample 3 (triple QD stack grown with 120s GIT), (d) reference single uncapped QD sample with related diameter histogram, and (e) uncapped triple QD sample grown with long GIT as sample 3 with related diameter histogram.
16

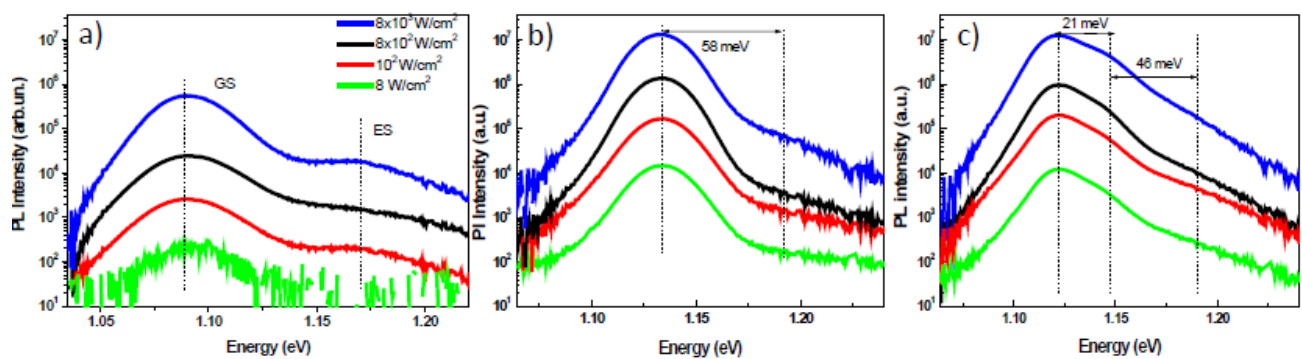

Figure 3: Backscattering PL spectra carried out at low temperature (T=10K) as a function of excitation power densities for (a) sample 1, (b) sample 2, and (c) sample 3.

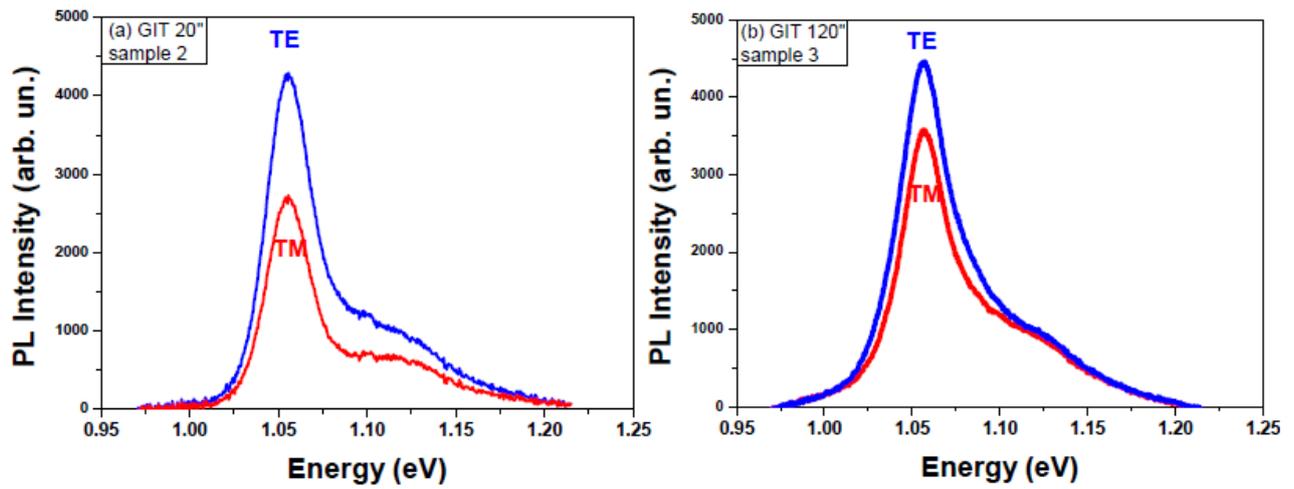

Figure 4: Polarized (TE and TM) room temperature PL measurements for the two trilayer QD samples under the same experimental conditions: (a) sample 2 grown with the GIT = 20s and (b) sample 3 grown with the GIT 120s. A significant increase in the TM/TE ratio is measured for the sample 3 with longer GIT, indicating a potential to tune polarization properties by controlling the growth dynamics.

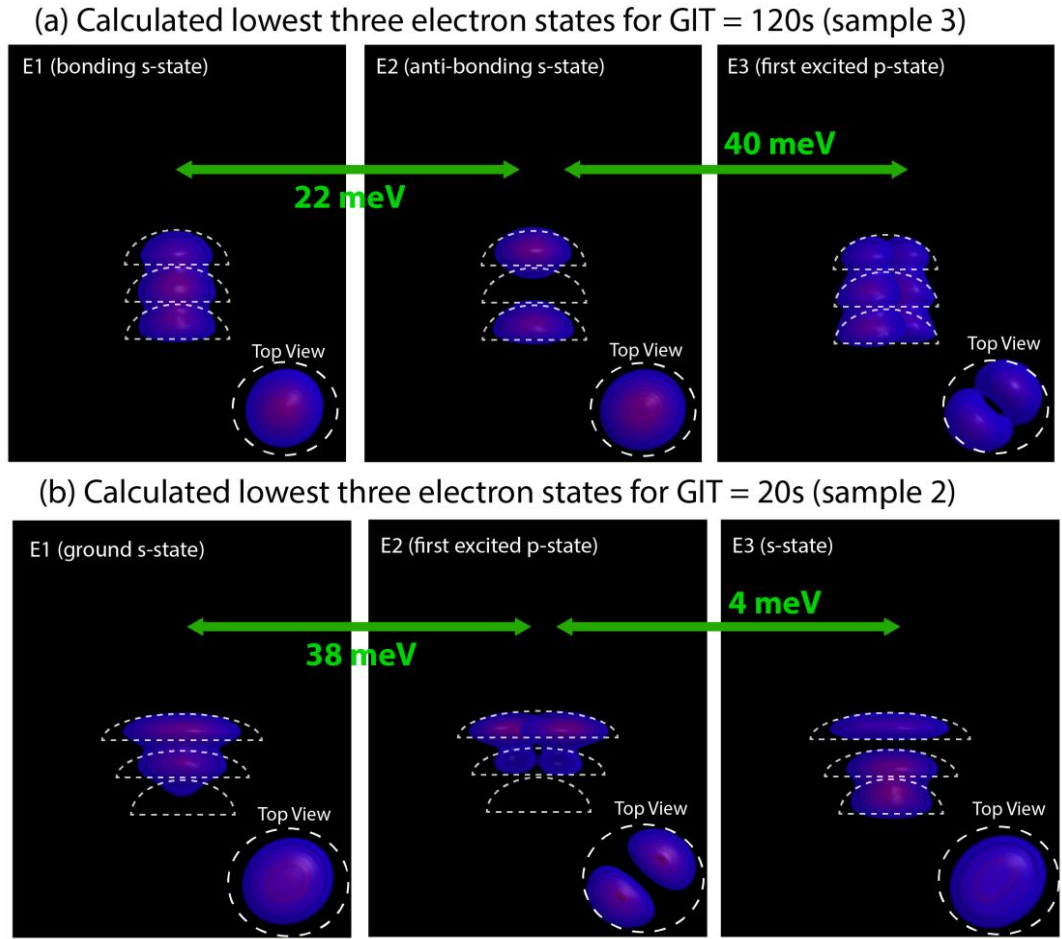

Figure 5: Plot of the lowest three conduction band states (E1, E2, and E3) calculated for the two model systems MS1 and MS2 corresponding to the two experimentally grown stacked samples, (a) sample 2 and (b) sample 3, respectively. In each case, we show side view of electron wave functions as well as the top view at the bottom right corner. The presence of strong bonding and anti-bonding s-states is clearly evident in (a) for the sample 3 with GIT=120s in accordance with the experimental PL measurements of figure 3(c). For the sample 2 grown with GIT=20s as shown in (b), the wave function hybridization effect becomes significantly weak, with the electron wave functions being largely present in the upper QDs of the stack with the larger size. Moreover the calculations do not exhibit the presence of bonding and anti-bonding states for this sample, again consistent with the experimental PL measurements of figure 3(b).

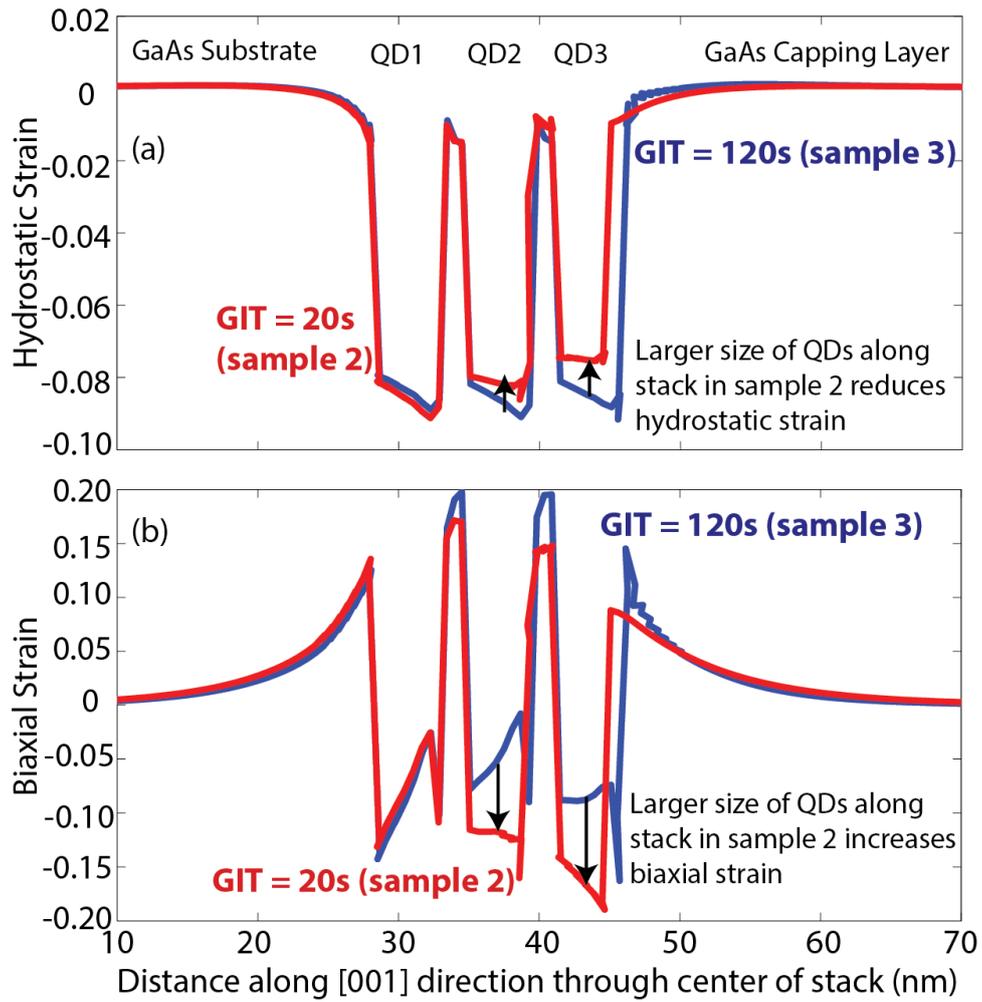

Figure 6: Hydrostatic and biaxial strain components are plotted in (a) and (b), respectively, for the two trilayer stack models, MS1 (sample 2) and MS2 (sample 3). Due to increasing size of QD layers in the sample 2, the hydrostatic strain reduces and the biaxial strain increases towards the top of the stack.

Table 1: The calculated values of the ground state peak energy (GS) and TM/TE ratio are provided for various double composition models (as shown by schematic diagram of figure 7) mimicking In/Ga intermixing effects in the sample 3. The GS and TM/TE values that are closest to the experimental values of GS ≈ 1.057eV and TM/TE ≈ 0.8 are highlighted by using bold fonts.

| In Composition Configuration # | Outer shell compositions | | | Inner core compositions | | | GS (eV) | TM/TE ratio |
|---|---|---|---|---|---|---|---|---|
| | x2 | x4 | x6 | x1 | x3 | x5 | | |
| Double Composition 1 | | 0.4 | | | 1.0 | | 0.98 | 1.047 |
| Double Composition 2 | | 0.4 | | | 0.9 | | 1.014 | 0.96 |
| Double Composition 3 | | 0.4 | | | 0.8 | | 1.07 | 0.88 |
| Double Composition 4 | | 0.3 | | | 1.0 | | 0.97 | 1.329 |
| Double Composition 5 | | 0.3 | | | 0.9 | | 1.02 | 1.218 |
| Double Composition 6 | | 0.2 | | | 0.8 | | 1.08 | 1.046 |
| Double Composition 7 | 0.4 | 0.3 | 0.2 | | 1.0 | | 0.98 | 0.92 |
| Double Composition 8 | 0.4 | 0.3 | 0.2 | | 0.9 | | 1.03 | 0.854 |
| **Double Composition 9** | **0.4** | **0.3** | **0.2** | | **0.85** | | **1.05** | **0.82** |
| Double Composition 10 | 0.4 | 0.3 | 0.2 | | 0.8 | | 1.07 | 0.76 |
| Double Composition 11 | 0.4 | 0.3 | 0.2 | 0.9 | 0.95 | 1.0 | 1.02 | 0.97 |
| Double Composition 12 | 0.4 | 0.3 | 0.2 | 0.85 | 0.9 | 0.95 | 1.04 | 0.938 |
| **Double Composition 13** | **0.4** | **0.3** | **0.2** | **0.8** | **0.85** | **0.9** | **1.06** | **0.88** |
| **Double Composition 14** | **0.4** | **0.3** | **0.2** | **0.75** | **0.8** | **0.85** | **1.08** | **0.83** |
| Double Composition 15 | 0.4 | 0.3 | 0.2 | 0.8 | 0.9 | 1.0 | 1.06 | 0.97 |

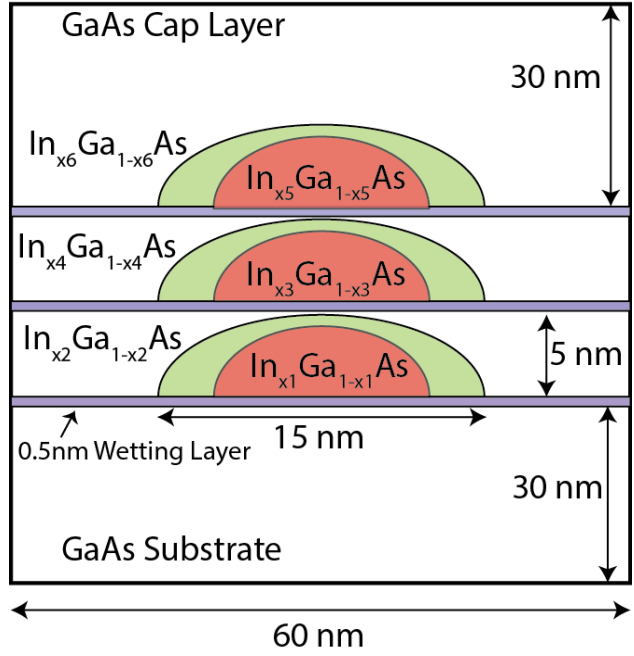

Figure 7: Schematic diagram of the trilayer quantum dot stack. Each quantum dot layer is comprised of a dome-shaped QD placed on top of 0.5 nm thick InAs wetting layer. The composition profile of the QDs is chosen according to the two-composition model to mimic the In-Ga interdiffusion effect, where the inner core (red region) is of higher In composition ($\geq 0.8$) and the outer shell (green region) is of lower In composition ($\leq 0.4$). The size of inner core region (red region) is 11 nm is diameter and 4 nm in height and is same for all the three QDs. The QD stack is embedded inside a very large GaAs buffer consisting of roughly 18 million atoms.

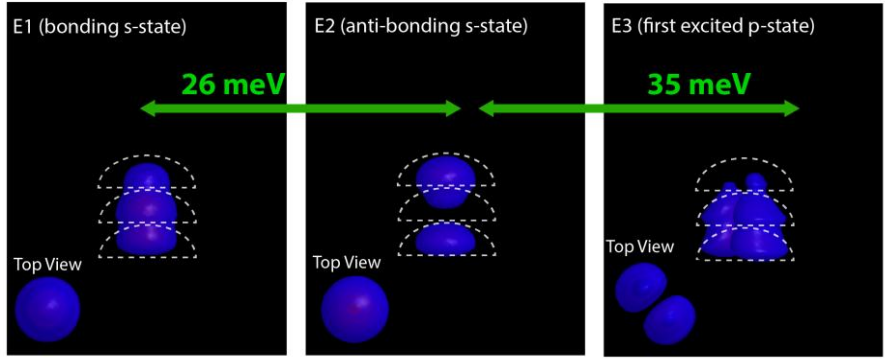

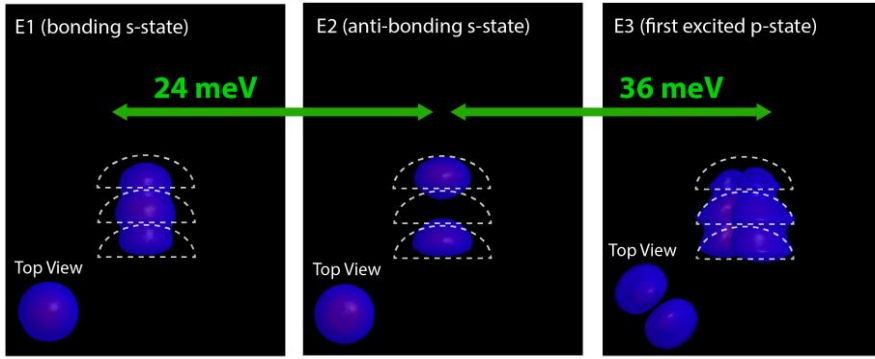

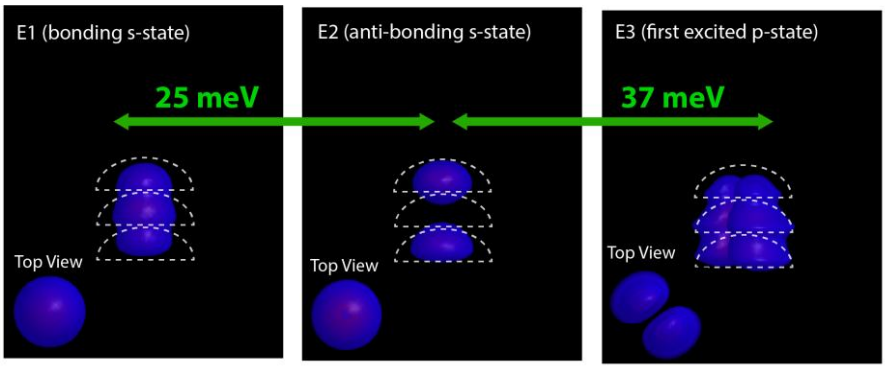

Figure 8: Plots of the lowest three conduction band states (E1, E2, and E3) calculated for the double-composition models 9, 13, and 14 (table 1). In each case, we show side views of the electron wave functions as well as the top views on the bottom left corners. The presence of strong bonding and anti-bonding states is clearly evident in all three cases, which conforms to the PL measurements performed on sample 3 (figure 3(c)).

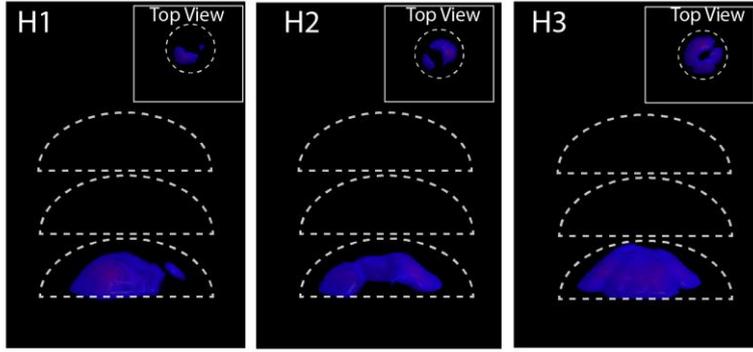
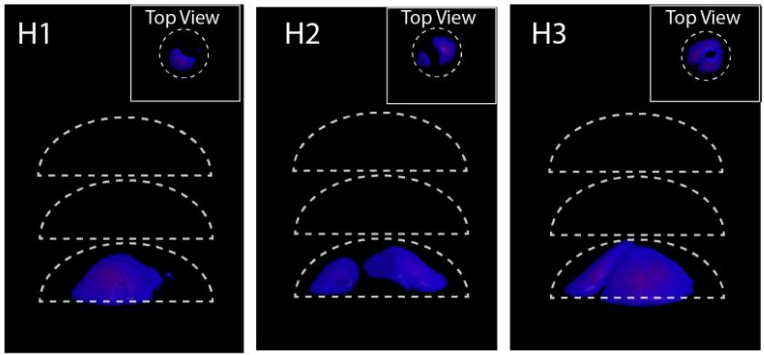
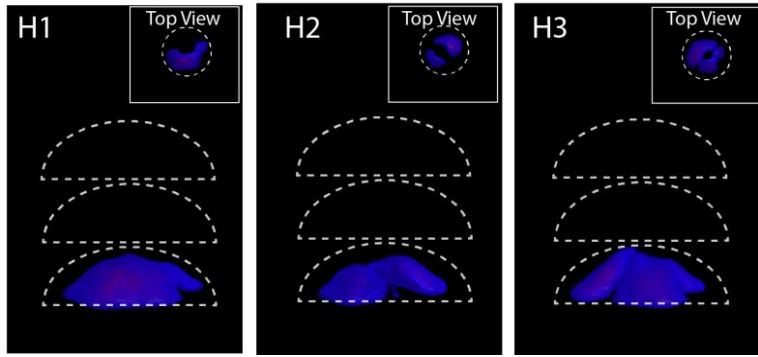

Figure 9: Plots of the highest three valence band states (H1, H2, and H3) calculated for the double-composition models 9, 13, and 14 (see table 1). In each case, we show side views of the hole wave functions as well as the top views on the top right corners.